\documentclass[prl,preprint,amsmath]{revtex4}
\input {epsf.sty}
\usepackage{epsfig}
\usepackage{amssymb}
\usepackage{natbib}
\bibpunct{}{}{,}{s}{}{}
\usepackage{setspace}

\begin{document}
\title{Discovery of an Excited Pair State in Superfluid $^3$He}
\author{J.P. Davis, J. Pollanen, H. Choi, J.A. Sauls, and W.P. Halperin}
\affiliation{Department of Physics and Astronomy, \\Northwestern 
University, Evanston, IL 60208,
USA}
\date{Version \today}
%\pacs{67.57.Jj, 67.57.-z, 43.35.Lq, 74.20.Rp}
%----------------------------------------------------------------------------------------------------

\begin{abstract} Order parameter collective modes are the fingerprint 
of a condensed phase. The
spectroscopy of these modes in superfluid $^3$He and unconventional 
superconductors can provide key
information on the symmetry of the condensate as well as the 
microscopic pairing mechanism
responsible for the ground state and excitation energies. We report 
the discovery of a new
collective mode in superfluid
$^3$He-B which we identify as an excited bound state of Cooper pairs. 
We use interferometry within
an acoustic cavity that is very  sensitive to changes in the velocity 
of transverse sound. Our
measurements of sound velocity and  mode frequency, together with the 
observation of acoustic
birefringence indicate that this new  mode is  weakly bound with an 
excitation energy within 1\% of
the pair-breaking edge of $2\Delta$. Based on the selection rules for 
coupling of transverse sound
to a  collective mode in $^3$He-B, combined with the observation of 
acoustic birefringence near the
collective mode frequency, we infer that the  new mode is most likely 
a spin-triplet ($S=1$),
$f$-wave pair exciton ($L=3$) with total angular momentum, $J=4$. The 
existence of a pair exciton
with
$J=4$ suggests an attractive, sub-dominant,
$f$-wave pairing interaction in liquid $^3$He.
\end{abstract}

\maketitle
%----------------------------------------------------------------------------------------------------
Fifty years ago Bardeen, Cooper and Schrieffer (BCS) published their 
seminal paper on  the theory
of superconductivity in metals \cite{Bar57}. This theory,  combined 
with  developments over the next
decade \cite{And58a,Gor58,Abr59b,Eil68}, is one of the  most 
successful theoretical achievements of
modern physics. The basic feature of BCS theory, the 
\emph{condensation} of bound  pairs of
fermions, has impacted research  on nuclear structure \cite{Boh58}, 
our  understanding of neutron
star interiors, the rotational dynamics of pulsars \cite{Pin85}, and 
most  recently, the physics of
ultra-cold, meta-stable phases of  atomic gases \cite{Zwi05,Alt07}. 
In condensed matter, BCS pairing
is ubiquitous. It is observed in  metals, magnetic materials 
\cite{Har05},  organic conductors
\cite{Jer82}, strongly  disordered films that are on the verge of 
becoming  insulators \cite{Mar99},
and liquid
$^3$He \cite{Leg75}.

Perhaps the most detailed and specific signatures of broken 
symmetries of the normal state are the
collective  modes of the pair condensate. These are the dynamical 
fingerprints of a multi-component
order  parameter \cite{Wol99}. The order parameter collective modes 
of  superfluid  $^3$He
  have been extensively studied \cite{Hal90} using acoustic absorption 
spectroscopy
\cite{Wol77,Lin87,Mov88,Hal90,Sau90,Ash96,Dav06,Dav07,Sau86}. There 
have been efforts to observe and
calculate the spectrum and  signatures of such modes in the  heavy 
fermion superconductors, UPt$_3$
\cite{Bat85,Hir92},  UBe$_{13}$ \cite{Gol85,Fel02}, as well as for 
Sr$_{2}$RuO$_{4}$ \cite{Fay00,
Miu07}.  Recently, observation of the Leggett-mode, i.e. the 
inter-band Josephson oscillations of a
two-band superconductor, was reported in MgB$_2$ \cite{Blu07}. 
Collective modes of the {\sl
amplitude} of the order parameter have also been observed.  In 
NbSe$_2$ the superconducting state
develops below the onset of a  charge density wave (CDW) instability. 
The coupling between the
amplitude mode of the CDW and the amplitude  mode of the 
superconducting order parameter produces an
infrared active collective mode  near the gap edge, $2\Delta$ 
\cite{Lit81,Soo80}.  However,
$^3$He is the best example  of complex symmetry breaking among the 
BCS condensates and investigation
of its order  parameter collective modes holds significance for the 
understanding of other pairing
systems.

Transverse sound has been shown to be a precision spectroscopy of the 
collective mode spectrum of
superfluid $^3$He-B \cite{Dav06,Dav07}. Using these techniquese we 
have discovered a new
collective mode  of the B-phase of superfluid $^3$He .  Selection 
rules for the coupling of
transverse sound to a collective mode, combined with the observation 
of acoustic birefringence near
the mode frequency, suggest that the  most likely candidate for this 
new  mode is the spin-triplet
($S=1$),
$f$-wave pair exciton ($L=3$) with total  angular momentum, $J=4$, 
predicted by Sauls and Serene
\cite{Sau81}.

The equilibrium phase of superfluid $^3$He-B is a spin-triplet 
($S=1$), $p$-wave  ($L=1$)
condensate.  The order parameter has nine complex amplitudes, and 
correspondingly a spectrum of
collective  modes \cite{Wol77,Hal90} whose observation has been 
instrumental in establishing that
the dominant pairing  interaction is $p$-wave. For pure $p$-wave 
pairing there are a total of 18
modes corresponding to the number  of degrees of freedom of the 
order parameter. The B-phase has an
isotropic gap with magnitude $2\Delta$, where
$\Delta/\hbar$ at zero temperature  varies from $34$ to $97$ MHz over 
the pressure range of  the
phase diagram. Acoustic techniques are well-suited  for these 
frequencies and, in general, sound
can couple effectively to order parameter modes with non-zero 
frequency  at zero wavevector,
$\Omega(k=0)\sim\Delta/\hbar$. These `optical' modes, shown in 
Fig.~1,  correspond to excited pair
states with total angular momentum quantum numbers \cite{Mak76},
$J= 0, 1, 2$,
\textit{etc}.,  each with $2J+1$ substates, labeled by $-J\leq m_{J}
\leq J$. Additionally, the order parameter modes are  classified by 
their parity under
particle-hole symmetry, $J^{\pm}$, where plus and minus distinguish 
between modes with even ($+$) or
odd  ($-$) parity  under particle-hole transformation, \textit{i.e.} particle
$\leftrightarrow$ hole conversion. In Fig.~1 we sketch  the energy 
level diagram for the excited
pair  states in superfluid $^3$He-B that have been observed, or are 
predicted, to couple to sound
\cite{Hal90,Wol77}.

\begin{figure}[t]
%--------------------- Figure 1 -----------------------
\centerline{\includegraphics[width=3.5in]{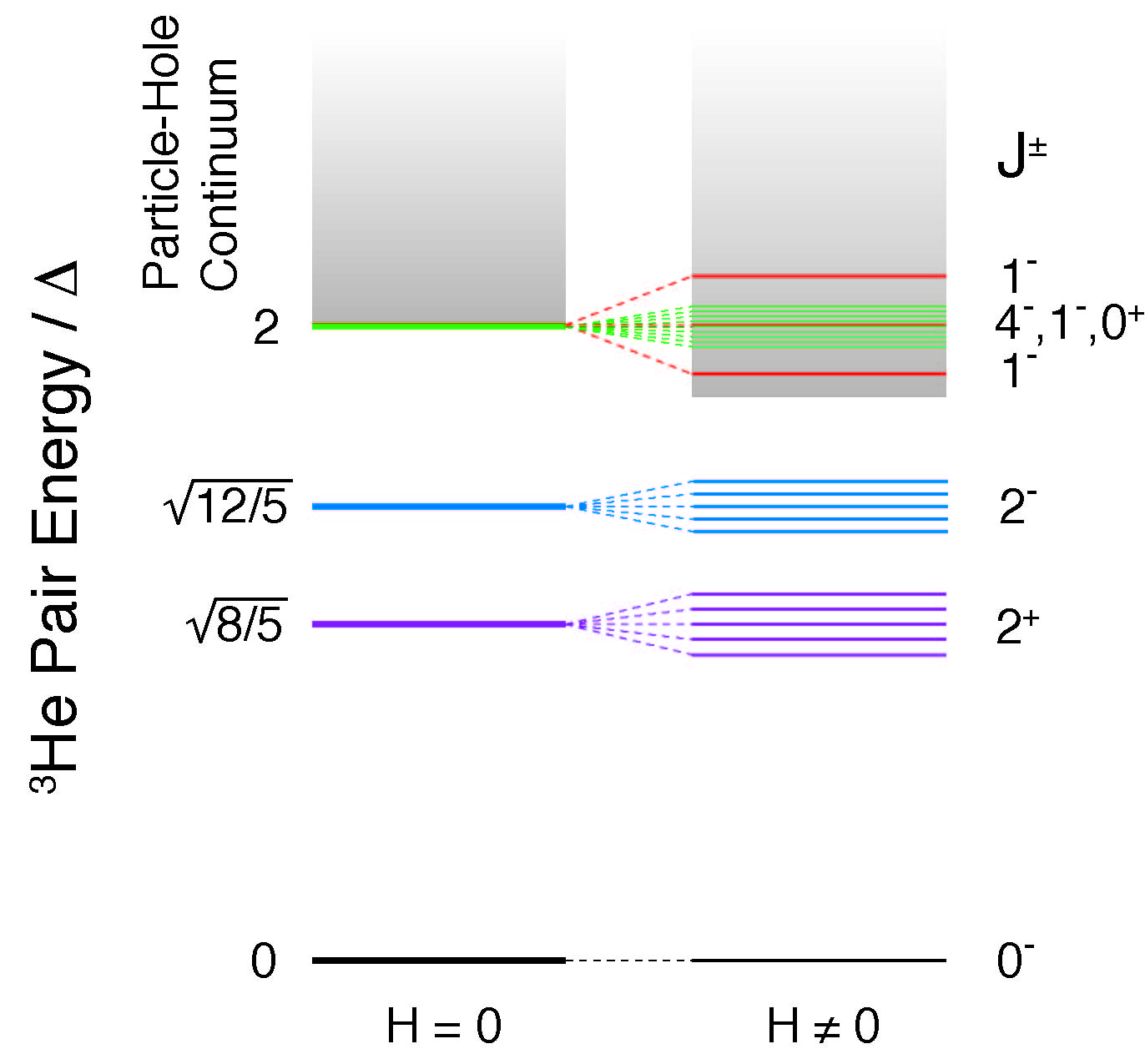}}
%------------------------------------------------------
\caption{\label{fig1}Energy levels of the collective  modes,
$J^{\pm}$, of superfluid $^3$He-B that have been observed or 
predicted to couple to either
longitudinal or transverse sound and their Zeeman splitting in a 
magnetic field in the limit  of
weak quasiparticle interactions. The $4^-$ and $0^+$ modes have not 
been observed and evidence  for
the $1^-$ mode is not yet well-established \cite{Ash96}.}
\end{figure}

Acoustic waves have a linear dispersion, $\omega = c\,k$, where $c$ 
is the sound velocity. The
sound frequency,
$\omega$, can be chosen to match an order parameter collective mode,
$\Omega_{J^{\pm},m_J}(T, P)$, which is  only weakly dependent on 
wavevector $k$, but can be  tuned
by sweeping the pressure or temperature.  If there is coupling 
between sound and the  modes they can be identified by divergence in the attenuation and velocity  of 
the propagating sound wave at
the crossing point
\cite{Hal90}. We use a unique spectroscopy based on 
magneto-acoustics that has high spectral
resolution \cite{Dav06,Dav07} and well-defined selection rules for 
coupling to the order parameter.
We constructed a cavity defined by an $AC$-cut quartz  piezoelectric 
transducer and a polished
quartz reflector that are separated by $D= 31.6 \pm  0.1~\mu$m; 
details are given in Sec. I of the
supplementary information. Transverse sound is both generated and 
detected by our transducer.  We
sweep the $^3$He pressure, holding the temperature  near $T\approx 550\,\mu$K
$\ll T_{c}$. The resulting  changes in  the phase velocity alter the 
number of half-wavelengths in
the cavity and are manifested  as acoustic  interference producing 
an oscillatory electrical
response, shown in Fig.~2A. The acoustic signal can  be represented 
as $ A = A_{0} + A_{1}
\cos\theta\sin(2D\omega/c + \phi)$,  where $A_{0}$ is a smoothly 
varying background in the absence
of cavity  wave-interference and is determined by acoustic  impedance 
\cite{Nag07};
$A_{1}$ is the amplitude of the oscillatory signal modulation from 
wave-interference; $\phi$ is
a phase angle; and $\theta$  is the angle between the polarization of 
the  sound wave at the surface
of the transducer and the direction of  linear polarization that the 
transducer can generate and
detect. We apply magnetic  fields, $H$, up to 305 G,  parallel to the 
propagation direction,
revealing acoustic birefringence where $\theta \propto H$. These 
acoustic  techniques offer a
precise means to investigate the order parameter structure of 
superfluid $^3$He.

Transverse sound does not ordinarily propagate in fluids. However, it 
was predicted by Moores and
Sauls  (MS) \cite{Moo93} to exist in superfluid $^3$He-B as a 
consequence of coupling to the
$J=2^{-}$, $m_{J} =\pm 1$  modes. The experiments of Lee {\it et al.}
\cite{Lee99} confirmed this theory which was later exploited by Davis 
{\it et al.}
\cite{Dav06,Dav07} to investigate the $J=2^{-}$ mode with much higher 
precision than is possible
with  longitudinal sound
\cite{Mov88}.  Furthermore, MS predicted that $^3$He-B becomes 
circularly birefringent in a
magnetic field, where the $J=2^-$, $m_{J} =\pm 1$ modes couple 
differently to right and  left
circularly polarized  transverse sound waves \cite{Moo93}, giving 
them different velocities.
Circular birefringence leads to rotation  of the plane of 
polarization of linearly polarized
transverse sound by an angle proportional to the component of 
magnetic field in the propagation
direction \cite{Moo93,Lee99,Dav07}, the acoustic analog of optical 
Faraday  rotation. Selection
rules govern the coupling of the collective modes with quantum 
numbers $J^{\pm},m_J$ to right- and
left circularly polarized transverse sound. Circularly polarized 
transverse waves in
$^3$He-B, propagating in the direction of the magnetic field, 
preserve axial symmetry. As a result,
transverse sound couples only to $m_J=\pm1$ modes.  Application of a 
magnetic field lifts the
degeneracy of the $m_J=\pm1$ states via the nuclear Zeeman energy, 
producing circular birefringence.
Therefore an  order parameter collective mode,
$\Omega_{J^{\pm},m_J}$, that induces acoustic circular birefringence requires
$J\geq1$ with $m_J=\pm1$. Additionally, MS have shown that in zero 
field only \emph{even} angular
momentum modes  with $m_{J}=\pm 1$ couple to transverse  currents 
\cite{Moo93}. These selection
rules follow from the invariance  of the B-phase ground  state under 
joint spin and orbital
rotations (i.e. a $J=0$ ground state) and are applicable for  the 
geometry  of the acoustic
experiments described here.

\begin{figure}[t]
%--------------------- Figure 2 -----------------------
\centerline{\includegraphics[width=6in]{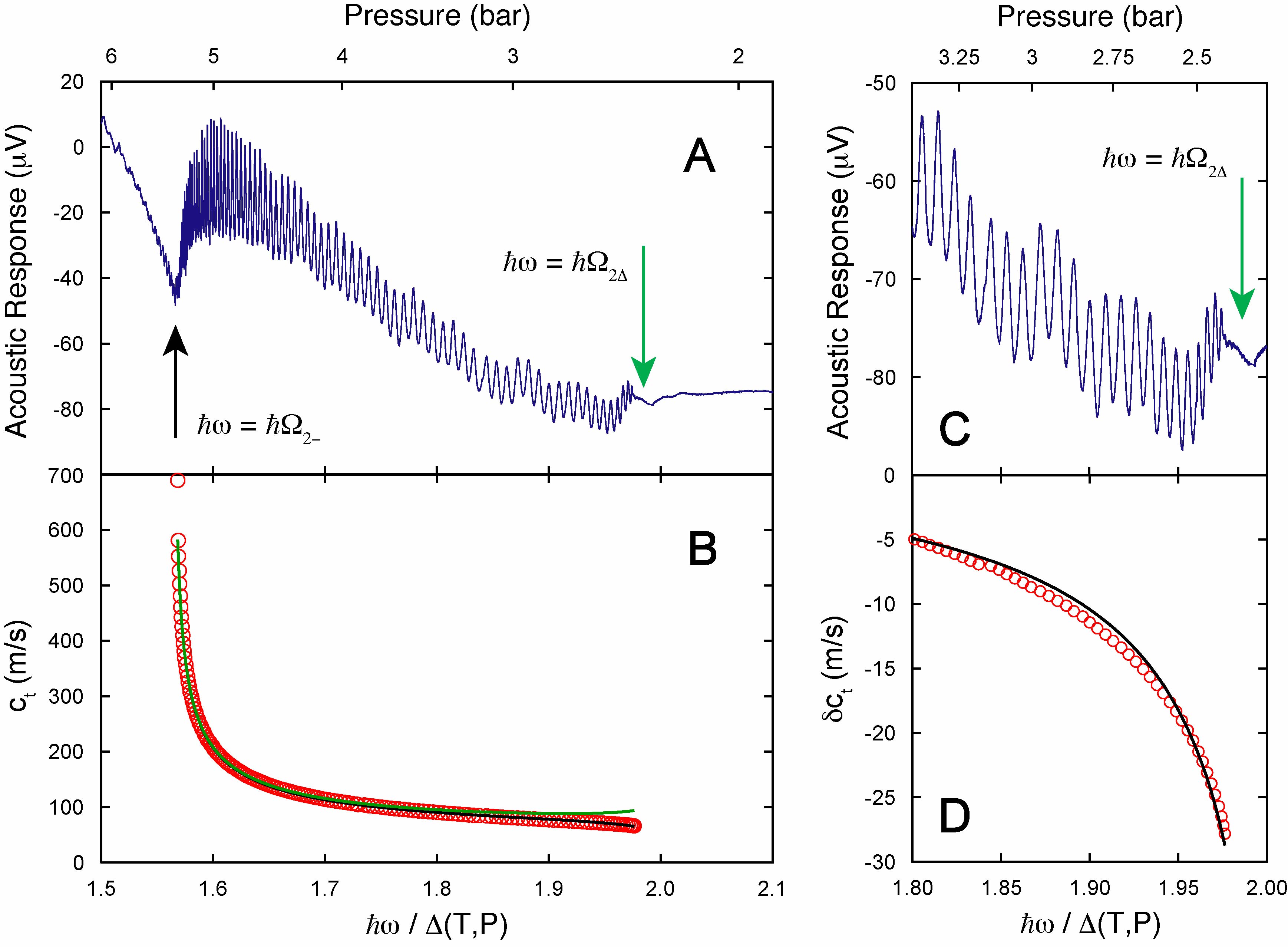}}
%------------------------------------------------------
\caption{\label{fig2}Acoustic cavity response to pressure as a 
function of energy
   	normalized to the gap energy, at $88$ MHz and $\approx 
550~\mu$K in zero magnetic field.
   	(\textbf{A}) Interference oscillations obtained from a 
pressure sweep, where the arrows
   	mark the well-established $J=2{^-}$ mode (black) and a new 
mode at the gap-edge (green).
   	(\textbf{B}) Transverse sound velocities (red circles) from 
(A) compared with the theory,
   	Eq.~1 (green curve).  The black curve is calculated by adding 
a  mode to Eq.~1. (\textbf{C})
   	Detail of (A) near $2\Delta$.  The green arrow marks where 
the period of the oscillations goes
   	to zero, indicating the $2\Delta$-mode,
$\hbar\Omega_{2\Delta}$. (\textbf{D}) The difference between the 
measured and theoretical (Eq.~1)
transverse sound velocities (red circles) and the difference between 
the phenomenological model and
Eq.~1 (black curve).}
\end{figure}

In the superfluid state, transverse sound propagates according to the 
dispersion relation
\cite{Moo93},
\begin{equation}\label{dispersion}
           \left(\frac{\omega}{q v_F}\right)^{2}_{m_J} =
	 \Lambda_{0} +
           \Lambda_{2^{-}}\frac{\omega^{2}}{\omega^{2} -
\Omega_{2^{-},m_J}(H)^{2}-\frac{2}{5}q^2 v_{F}^{2}}.
\end{equation} This equation can be solved for the phase velocity as 
shown by the green curve in
Fig.~2B, where $v_F$  is the Fermi velocity,
$\Lambda_0$ is the quasiparticle  restoring force, and the $2^{-}$ 
mode frequency is
$\Omega_{2^-,m_J}(H) = \Omega_{2^-}+m_{J}g_{2^{-}}\gamma_{eff} H$, 
to first order in magnetic
field. The Land\'e $g$-factor, $g_{2^{-}}$, gives the Zeeman 
splitting \cite{Dav07}  of the mode,
$\gamma_{eff}$ is the effective gyromagnetic ratio \cite{Sau82}  of 
$^3$He, and the complex
wavevector is $q=k+i\alpha$ where $\alpha$ is the attenuation. For 
substates $m_J=\pm 1$, there is
a non-zero magneto-acoustic coupling, $\Lambda_{2^-}$, between the
${2^-}$  mode and transverse sound. Further details for $\Omega_{2^{-},m_J}$,
$\Lambda_{0}$ and $\Lambda_{2^-}$ are given in Sec. II of the 
supplementary information.

  From Eq.~1 it is apparent that the sound velocity diverges for 
acoustic frequencies approaching
the  ${2^-}$ mode,  resulting in faster oscillations in the acoustic 
response as is evident in
Fig.~2A. Quantitative comparison with theory can be made by 
converting the oscillations into the
transverse sound phase velocity. In order to obtain absolute  values 
for the velocity we fix one
adjustable parameter by comparison of our data with the velocity 
calculated  from Eq.~1 near the
${2^-}$ mode, as shown in Fig.~2B. The agreement between the 
calculation, green curve, and the
experiment, red circles, is excellent for energies below
$\sim 1.8\Delta$. However, above this energy we observe a downturn in 
the transverse sound
velocity.   This effect is quite clear in the raw data displayed in 
Fig.~2C indicated by a
decreasing period of the  oscillations with increasing energy. This 
is a classic signature of the
approach to an order parameter collective  mode \cite{Hal90}. In 
Fig.~2D we show as red circles the
difference between the measured sound velocity and the  value 
calculated from Eq.~1, based on
coupling only to the ${2^-}$ mode.  The downturn in  velocity is 
quite evident here.  By
extrapolating the period of the oscillations in Fig.~2C to zero, i.e. 
to the point where the
velocity diverges indicated by the green arrow, we determine the 
frequency  (excitation energy) of
the new collective mode. This procedure is described in detail in 
Sec. III  of the supplementary
information. We find the excitation energy of the mode, 
$\hbar\Omega_{2\Delta}$,  to vary
systematically with pressure, but  remain within 1\% of $2\Delta$ for 
pressures from 1 to  20 bar,
as shown in Fig.~3A. The precision of the  extrapolation is given by 
the error bars; the  accuracy
of 1\% is determined by the absolute temperature scale \cite{Gre86} 
in the framework of the
weak-coupling-plus model for the energy gap \cite{Rai76}.

In order to model the sound velocity near the gap-edge, we amend 
Eq.~1 by adding a new term to
represent the  coupling of transverse sound to the $2\Delta$ 
collective mode with a form similar to
that of the ${2^-}$ mode,
$\omega^{2}\Lambda_{2\Delta}/(\omega^{2} - \Omega_{2\Delta}^{2})$. 
The velocity calculated from
this model is  given by the black curve in Figs.~2B and 2D, 
describing our data quite  well with a
coupling strength,
$\Lambda_{2\Delta}= 0.18$.  Furthermore, we note that the amplitude,
$A_{1}$, of the interference oscillations near the gap-edge 
decreases in a manner  similar to the
period.  Since the amplitude is  proportional to the inverse of the 
sound  attenuation, this is a
consistent identification of a collective mode, where it can be 
shown from  Eq.~1  that the
attenuation diverges at the mode location, as does the velocity.

\begin{figure}[t]
%------------------------- Figure 3 ------------------------------------
\centerline{\includegraphics[width=7in]{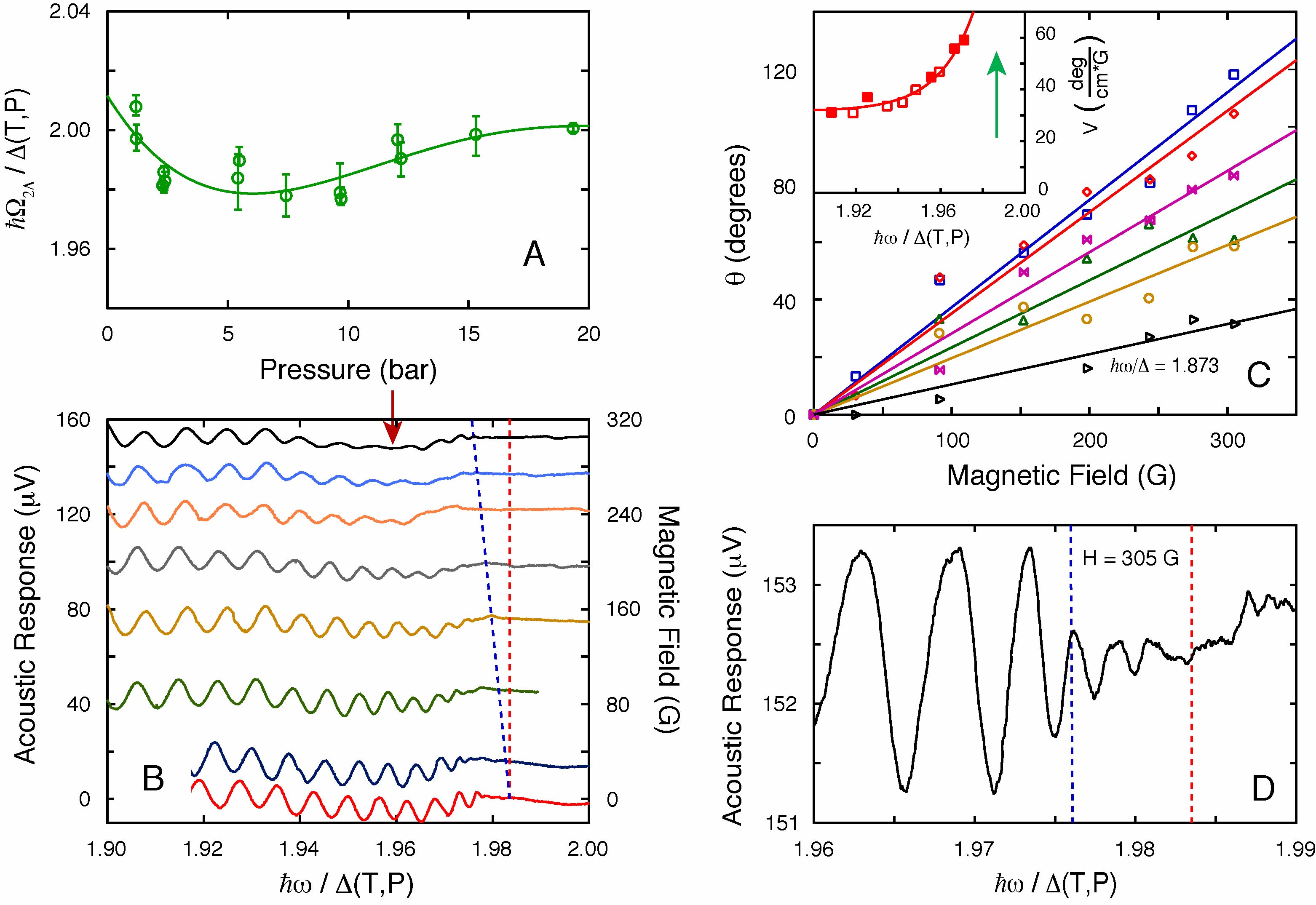}}
%-----------------------------------------------------------------------

\caption{\label{fig3}(\textbf{A}) $\hbar\Omega_{2\Delta}$ as a 
function of pressure in zero magnetic
field. The curve is a guide to the eye. (\textbf{B}) The acoustic 
response as a function of energy
offset by  magnetic field (right axis)	at $88$ MHz and $\approx 
550~\mu$K.  Acoustic  birefringence
rotates the plane of polarization by an angle $\theta$, producing a 
minimum signal amplitude at
$\theta = \pi/2$, indicated by the red arrow for 305 G and $\approx 
1.96\Delta$. The blue dashed
line represents the predicted field dependence ($g=0.4$) of the 
$J=1^-$, $m_{J}=-1$ mode.
(\textbf{C}) The Faraday rotation angle, $\theta$, is proportional to 
magnetic field at energies
given by solid squares in the inset. Inset: Verdet constant as a 
function of energy, diverging at
$\hbar\Omega_{2\Delta}$, marked by the green arrow. (\textbf{D}) 
Detail of the 305 G trace from (B).}
\end{figure}

We have previously established \cite{Lee99} that an applied magnetic 
field rotates the plane of
polarization of  propagating transverse sound in the near vicinity of 
the $J=2^{-}$ mode, $\omega
\geq \Omega_{2^{-}}$.  Increasing the frequency above the
$J=2^{-}$ mode decreases the Faraday  rotation rate, which eventually 
becomes immeasurably small  as
the coupling to the mode decreases.  However, at even higher 
frequencies near the  pair-breaking
edge we find that the Faraday rotation  reappears (see Fig.~3B) in 
the same frequency region where
we observe the downturn in the velocity  of transverse sound. The 
magnetic field modulates the
interference amplitude,
$A_{1}$, by a factor
$\cos\theta$, from which we extract the Faraday rotation angle,
$\theta$. In magneto-optics, Faraday  rotation is parameterized by 
the Verdet constant,
$V=\theta/2DH$. For our magneto-acoustic data we find
$V$ to be a monotonically increasing function of frequency, diverging 
near $\Omega_{2\Delta}$,
shown  as a green arrow in the inset of Fig.~3C.  Aparently the 
birefringence  originates from the
$2\Delta$  collective mode. The existence of acoustic  Faraday 
rotation (circular birefringence)
requires that the $2\Delta$-mode correspond to an  excitation with 
total angular momentum
$J\geq1$ with substates $m_{J} =\pm 1$ coupling to transverse 
currents. It is particularly
noteworthy that  the divergences in velocity and attenuation occur in 
both zero and  non-zero
applied fields at the excitation energy shown in Fig.~3A.

Order parameter modes predicted to lie close in energy to the 
pair-breaking edge include the
$J=0^{+}, 1^{-}, 4^{-}$ modes (see Fig.~1).
   Observation of the $0^{+}$ and $4^{-}$ modes have not previously 
been reported .
   The observation of circular birefringence allows us to rule out the
$0^{+}$ mode.
   A peak in the longitudinal sound attenuation that appears for 
non-zero magnetic  field
\cite{Lin87,Sau90} was attributed \cite{Ash96,Mck93} to the coupling 
of the $J=1^{-}, m_{J}=0$ mode
to the longitudinal current (see Sec. IV of the  supplementary information).
   However, in zero field the $J=1^-$ modes do not contribute to the 
stress tensor and cannot couple
to either the longitudinal or transverse currents. Consequently, 
neither the $J=1^-$ nor the $J=0^+$
modes can account  for our observations in zero field.
   Furthermore, in a non-zero  magnetic field the predicted Zeeman 
splitting \cite{Sch87} of the
$J=1^-$ modes should decrease the energy of the
$m_{J}= -1$ level as shown by the blue dashed line in  Fig.~3D, 
overlapping a region of energy
where we have  observed acoustic interference oscillations.  Since 
the velocity is not perturbed in
this region, it appears that  the Zeeman splitting of the
$2\Delta$-mode is much smaller than that predicted for the $1^-$ mode and so the
$1^-$ mode  cannot explain our data in a magnetic field.
   Finally, pair breaking is absent for frequencies below a threshold 
$2\Delta$  in zero magnetic
field, and thus cannot account for the downturn in velocity below 
$2\Delta$. Higher angular momentum
modes, in particular modes with $J=4^-$,
$m_J=\pm 1$, do couple to transverse sound even in zero field. Based 
on the experimental geometry
and the selection rules for  transverse sound and acoustic 
birefringence, the only known theoretical
candidate that might account  for our observations are the $J=4^-$ 
modes.  Quantitative predictions
for the coupling strength  and the magnitude of the Zeeman splitting 
are needed for a conclusive
identification.

Predictions for the $J=4^-$ mode frequency \cite{Sau81} depend on 
existence of an attractive
$f$-wave pairing  interaction\cite{Sau86}, subdominant with respect 
to $p$-wave pairing, as well as
the unknown Fermi  liquid parameter, $F^{s}_{4}$.  Several 
experiments, including magnetic
susceptibility and $J=2^{\pm}$ collective mode  spectroscopy,
\cite{Sau82,Fis87,Fis88,Hal90,Dav06,Dav07}  were analyzed to try and 
determine the $f$-wave pairing
interaction, as well as Fermi liquid interactions, in an effort to 
predict the $J=4^-$ mode
frequency. However, the results of  these different analyses are 
ambiguous owing to imprecision of
the Fermi liquid parameters, $F^{a,s}_{2}$,  as well as non-trivial 
strong  coupling effects
\cite{Dav07}.
   Our observation of a new order parameter collective mode, 
identified  on the basis of selection
rules as the
$J=4^{-}$ mode, provides direct evidence for an attractive $f$-wave 
pairing interaction. Precise
measurements of the mode frequency in this work, combined with a 
future  measurement of the Zeeman
splitting, should provide constraints on the microscopic pairing 
mechanism in $^3$He. Additionally,
our results may lead to realization of predictions of mixed symmetry 
pairing near impurities,
surfaces, and interfaces; surface phases with broken time-inversion; 
and novel vortex phases
\cite{Bal06,Fog97}.

We have discovered an excited pair state in superfluid $^3$He-B near
$2\Delta$ from measurements of the  divergence of velocity, 
attenuation, and the magneto-acoustic
Verdet constant for transverse sound.  Together with selection rules 
for acoustic  birefringence,
our experiments indicate that the $2\Delta$-mode  has total angular 
momentum $J\ge 4$
   and a frequency within 1\% of $2\Delta$ at all pressures. The 
observed mode is most likely the
$J=4^-$ mode predicted by Sauls and Serene \cite{Sau82} and, as such, 
provides direct evidence for
$f$-wave pairing correlations in superfluid $^3$He. Our results give 
a more detailed picture of the
order parameter structure in superfluid $^3$He-B near the  gap-edge 
than has been possible with
longitudinal sound and will serve as a guide to future work on  this 
and other unconventionally
paired systems.

We acknowledge support from the National Science Foundation, 
DMR-0703656 and thank  W.J. Gannon,
M.J. Graf, Y. Lee, M.W. Meisel, and B. Reddy for useful discussions.

\begin{singlespace}

\end{singlespace}
\clearpage
\section{Supplementary Information}

\section{I. Experimental details}

In our experimental arrangement, the cavity spacing of $D= 31.6 \pm 
0.1~\mu$m was measured {\it in
situ} with longitudinal sound at a temperature of 18 mK, using the
$17^{th}$ transducer harmonic which generates a large amplitude 
longitudinal signal.  We note that
the texture of the order parameter must be homogeneously  oriented 
along  the sound propagation and
magnetic field directions, since the diameter of the  transducer 
(0.84  cm) is much larger than the
spacing of the cavity. The frequency and  amplitude of the transducer 
resonances are extremely
sensitive to the acoustic conditions of the  cavity which modify the 
electrical  impedance thatwe
monitor with a continuous-wave,  RF-bridge \cite{Dav06,Dav07}.   We 
use odd harmonics of our
transducer from the 13$^{th}$ to the  29$^{th}$, corresponding to 
frequencies from 76 to 171 MHz.
Details of  cooling and thermometry  can be found elsewhere
\cite{Dav07}.  We use the weak-coupling-plus model for the gap
\cite{Rai76} as tabulated by Halperin and Varoquaux~\cite{Hal90}, 
locked to the  Greywall
temperature scale
\cite{Gre86}.  This allows us to account for variations of the 
transverse sound velocity with
changes in both pressure and temperature, although temperature 
effects are very slight at the low
temperatures used in this experiment. The accuracy of the  Greywall 
temperature scale
\cite{Gre86} determines the accuracy of $\Delta$ in the 
weak-coupling-plus model  and is estimated
to be 1\%.

\section{II. Calculation of the transverse sound velocity}

In order to calculate the transverse sound  velocity from Eq.~1, 
which appears in Figs.~2B and 2D,
we use the full expression for the
$J=2^-$ mode frequency in a magnetic field given by Moores and 
Sauls\cite{Moo93}
$\Omega_{2^-,m_J}(H)^2 =
\Omega_{2^-}^2 + 2m_{J}g_{2^{-}}\gamma_{eff} H\omega$, for
$m_{J}=\pm1$.  Additionally, we used the quasiparticle restoring 
force for transverse sound
\cite{Moo93} $\Lambda_{0} 
=\frac{F_1^{s}}{15}(1-\lambda)(1+\frac{F_{2}^{s}}{5})/(1+\lambda\frac{
F_{2}^{s}}{5})$, which includes all quasiparticle interaction terms 
\cite{Hal90},
$F_l^s$, for $l\leq2$. Similarly
$\Lambda_{2^{-}} =
\frac{2F_1^{s}}{75}\lambda(1+\frac{F_{2}^{s}}{5})^{2})/(1+\lambda\frac{F_{2}^{s}}{5})$. 
The Tsuneto
function $\lambda(\omega,T)$ can be thought of as a frequency 
dependent superfluid density
\cite{Moo93}, that depends on the gap amplitude,
$\Delta$.  Therefore, at low temperatures the quasiparticle term is 
small and $\Lambda_{2^{-}}$ is
enhanced.  In our calculations of  velocity we use
$\lambda(\omega,T)$ adapted to incorporate the weak-coupling-plus gap 
as described above
\cite{Rai76}.  The
\begin{figure}[t]
%%%%%%%%%%%%%%%%%   F I G U R E  SOM1  %%%%%%%%%%%%%%%%%%
        \centerline{\includegraphics[width=5in]{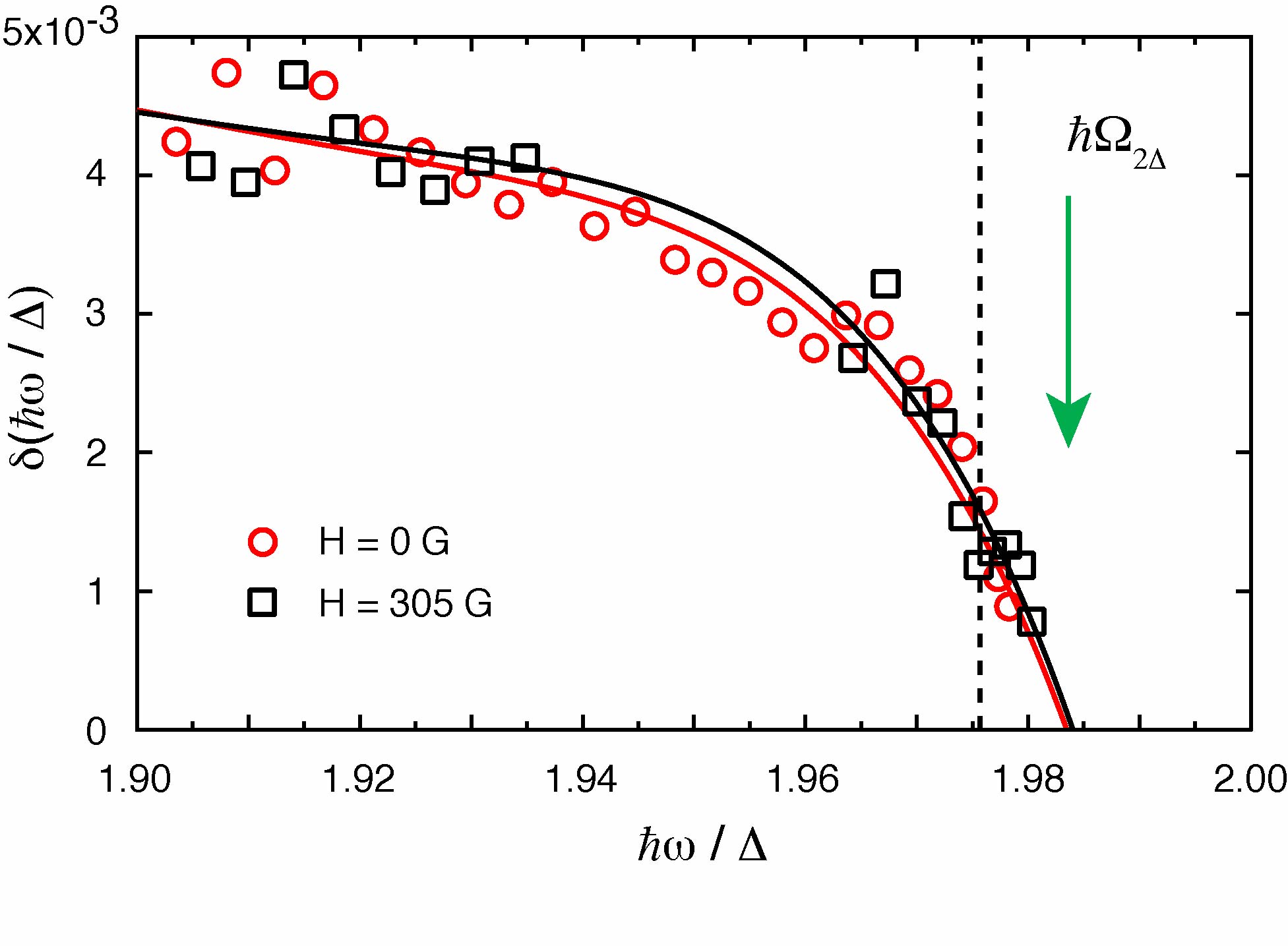}}
%%%%%%%%%%%%%%%%%%%%%%%%%%%%%%%%%%%%%%%%%%%
\label{SOM1}\vspace{-24pt}\end{figure}
\begin{singlespace}
\hspace{-0.25in}
\begin{minipage}{6.5in}{\small FIG.~S1. Period of acoustic energy 
oscillations from the acoustic
cavity interference of Fig.~3B as the acoustic frequency approaches  the
$2\Delta$ mode, green arrow, for zero magnetic field (red circles) 
and 305 G (black  squares).  We
identify
$\hbar\Omega_{2\Delta}$ from the extrapolation to zero (where the 
velocity diverges).  Both traces
point to the same energy within experimental resolution.  For 
comparison the energy  for the
collective mode $J=1^-$,
$m_J=-1$ in $H=305$ G with a Zeeman splitting of
$g=0.4$ is shown as a black  dashed line as expected theoretically
\cite{Sch87}.  The absence of data  over a small  range near 1.95 for 
the 305 G trace corresponds
to the region where Faraday rotation  has decreased the amplitude of 
the acoustic oscillations  and
the period is not reliably determined.}
\end{minipage}
\end{singlespace}

\vspace{24pt}
\noindent dispersion relation, Eq.~1, holds in the long  wavelength 
limit, $kv_F \ll \omega$.
Consequently, our estimation of the coupling strength of  the 
$2\Delta$-mode may be modified in a
full $q$-dependent analysis.
\section{III. Locating the new mode}

As the acoustic frequency approaches that of an order parameter 
collective mode, the sound velocity
and attenuation diverge.  In order to pinpoint the frequency  of the 
mode we plot the inverse of
the signal oscillation amplitude (attenuation) and the  oscillation 
period  (velocity) and
extrapolate to zero.  In Fig.~S1 we show this procedure for the 
oscillation period  for two
experiments with zero magnetic field and $H=305$ G.  The curves are 
fits to guide the eye.
Nonetheless it is clear that the mode frequency can be precisely 
determined.  This is the method
used to obtain Fig.~3A.  Moreover, the $2\Delta$ mode does not appear to have a
    Land\'e $g$-factor that is large enough to appear directly as a 
shift of the data trace in
$H=305$ G. As a point of comparison, if the Land\'e $g$-factor were 
to be of  the magnitude
theoretically predicted for the $J=1^-$ mode a shift in frequency 
indicated by the  dashed line
would be expected \cite{Sch87}.

\section{IV. Longitudinal sound and the $J=1^-$ mode}

Ling {\it et al.} \cite{Lin87} reported the existence of a mode, with 
a splitting which they
ascribed to $J=1^-$, $m_{J}=\pm1$.  These signatures in longitudinal 
sound attenuation appeared
only in the presence of an applied magnetic field, which in their 
case was oriented transverse to
the sound propagation direction. Since then it has been shown
\cite{Mck93,Ash96} that only the
$m_{J}= 0$ mode can couple to longitudinal sound, contrary to what 
was originally proposed
\cite{Lin87,Sch87}.  Additionally, Ashida {\it et  al.}
\cite{Ash96} find that the anomalies observed in longitudinal sound 
attenuation in non-zero field
\cite{Lin87} can be accounted for in terms of pair-breaking coupled 
to $J=1^-$, $m_{J}= 0$ and
$J=2^-$, $m_{J}=0,\pm2$ modes, in some combination, but that the 
analysis is complicated by the
inhomogeneous texture associated with the experimental conditions of 
a transverse magnetic field.
   From their analysis it appears that identification of the
$J=1^-$ mode is not yet well-established.  A comparable calculation 
has not been performed for
transverse  sound; however, on symmetry grounds the $J=1^-$ mode 
cannot couple to transverse sound
in zero magnetic field, and in a non-zero field the coupling to 
transverse sound is indirect via the
$J=2^-$ mode.  We estimate the coupling strength for this process to 
be very weak compared to what
we measure for the
$2\Delta$ mode providing additional evidence that the $J=1^-$ mode is 
not responsible for our
observations.
\end{document}